\documentclass[journal=jacsat,manuscript=article]{achemso}

\usepackage[version=3]{mhchem} 

\author{Božidar N. Šoškić}
\affiliation{Faculty of Natural Sciences and Mathematics, University of Montenegro, Džordža Vašingtona bb, 81000 Podgorica, Montenegro}
\alsoaffiliation{Department of Physics and NANOlab Center of Excellence, University of Antwerp, Groenenborgerlaan 171, B-2020 Antwerp, Belgium}
\author{Jonas Bekaert}
\affiliation{Department of Physics and NANOlab Center of Excellence, University of Antwerp, Groenenborgerlaan 171, B-2020 Antwerp, Belgium}
\email{jonas.bekaert@uantwerpen.be}
\author{Cem Sevik}
\affiliation{Department of Physics and NANOlab Center of Excellence, University of Antwerp, Groenenborgerlaan 171, B-2020 Antwerp, Belgium}
\author{Milorad V. Milošević}
\affiliation{Department of Physics and NANOlab Center of Excellence, University of Antwerp, Groenenborgerlaan 171, B-2020 Antwerp, Belgium}
\email{milorad.milosevic@uantwerpen.be}

\title[An \textsf{achemso} demo]
{Enhanced superconductivity of hydrogenated $\beta_{12}$ borophene}


\begin{document}







\begin{abstract}
\textbf{Borophene stands out among elemental two-dimensional materials due to its extraordinary physical properties, including structural polymorphism, strong anisotropy, metallicity, and the potential for phonon-mediated superconductivity. However, confirming superconductivity in borophene experimentally has been evasive to date, mainly due to the detrimental effects of metallic substrates and its susceptibility to oxidation. In this study, we present an \textit{ab initio} analysis of superconductivity in the experimentally synthesized hydrogenated $\beta_{12}$ borophene, which has been proven to be less prone to oxidation. Our findings demonstrate that hydrogenation significantly enhances both the stability and superconducting properties of $\beta_{12}$ borophene. Furthermore, we reveal that tensile strain and hole doping, achievable through various experimental methods, significantly enhance the critical temperature, reaching up to 29 K. These findings not only promote further fundamental research on superconducting borophene and its heterostructures, but also position hydrogenated borophene as a versatile platform for low-dimensional superconducting electronics.}

\begin{figure}[t!]
    \centering
    \includegraphics[width=0.7\textwidth]{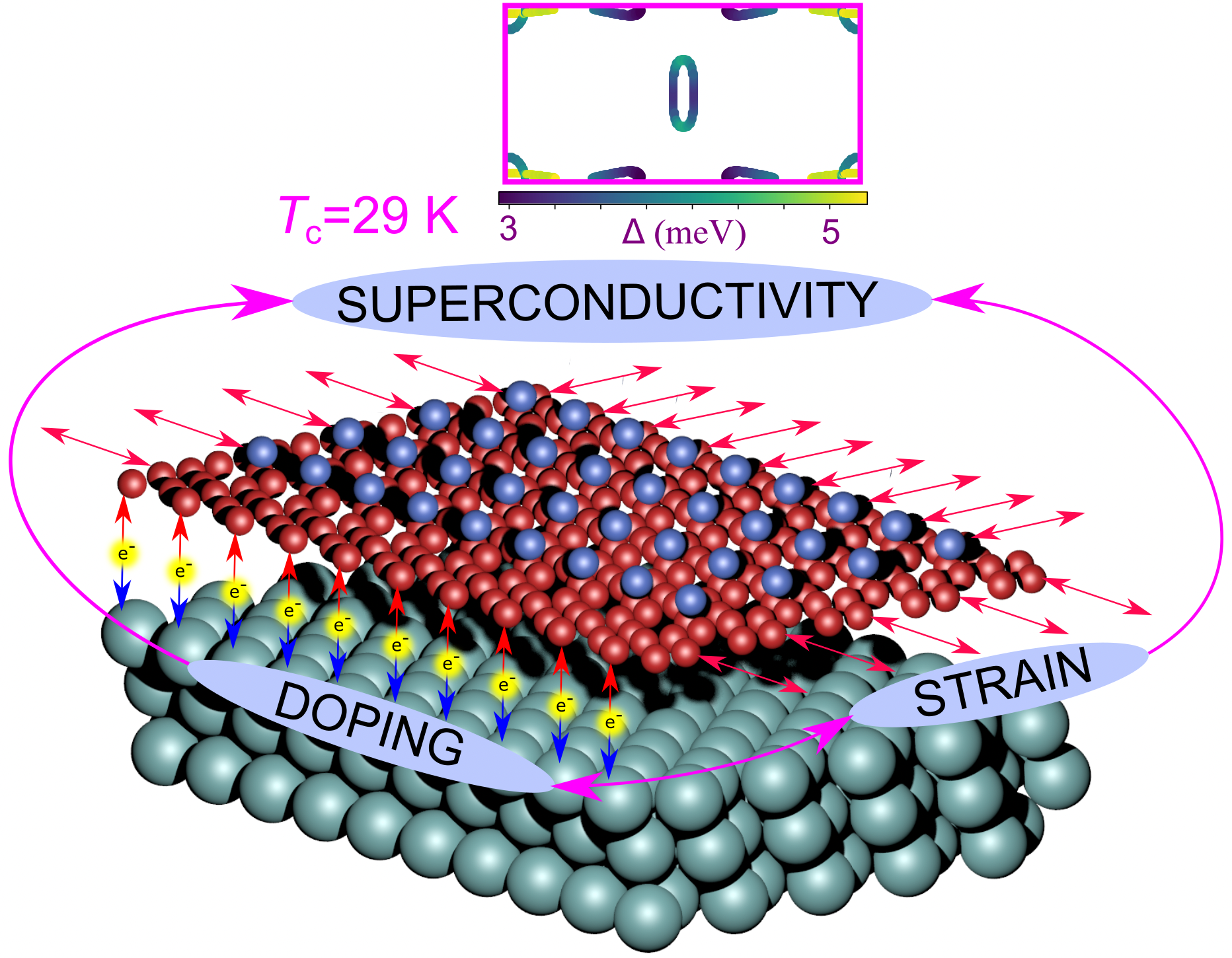}
    \label{fig:enter-label}
\end{figure}
  
\end{abstract}


With its intrinsic electron deficiency and more complex bonding nature than carbon \cite{Ogitsu}, boron ranks among the most chemically flexible elements, forming at least sixteen elemental bulk polymorphs \cite{Xu}, as well as a plethora of cluster structures of diverse sizes \cite{Oganov}, including (quasi-)planar forms \cite{Boustani}. On that front, theoretical predictions \cite{Yakobson1, Yakobson2} paved the way for the first experimental synthesis of two-dimensional (2D) boron structures, called borophenes, on a Ag(111) substrate in 2015 \cite{Manix, Feng}. These planar structures harbor a myriad of hexagonal voids with different concentrations (described by the $\eta$ parameter) and configurations \cite{Yakobson3}. The versatility of these borophenes is further demonstrated by their ability to be synthesized on a broad range of metallic substrates \cite{au,cu,ir_vinogradov}, transferred onto other, nonmetallic substrates \cite{transfer}, and ultimately isolated in freestanding form via liquid-phase exfoliation \cite{freestanding}.

Borophenes exhibit a plethora of apposite physical properties, including mechanical compliance~\cite{mp1,mp2}, high optical transparency \cite{ot}, high hydrogen storage capacity~\cite{hs1,hs2,hs3}, ultrahigh thermal conductance~\cite{tc}, the presence of metallic Dirac fermions \cite{dirac1,dirac2}, and possible hosting of phonon-mediated superconductivity \cite{sup1,sup2,PhysRevMaterials.8.064803}. Hence, the potential utility of borophenes spans various applications, from electrodes in batteries \cite{batteries} to catalysts for oxygen reduction \cite{oxygen} and detectors for gaseous substances~\cite{sensor1,sensor2}. However, practical applications may be hindered by borophene's susceptibility to chemical degradation, e.g. by oxidation, in ambient conditions \cite{ox1,ox2,ox3}. Surface hydrogenation emerges as a highly promising chemical passivation strategy, demonstrated to suppress oxidation of Si surfaces \cite{si_h} and 2D materials such as silicene \cite{silicene} and germanene \cite{germanane}. A significant advancement in that respect took place in 2021 with the synthesis of several hydrogenated $\beta_{12}$ ($\eta=\frac{1}{6}$) borophene structures (``borophanes'') on a Ag(111) substrate \cite{borophane}. Borophanes exhibit remarkable stability in ambient conditions, with borophene readily recoverable through thermal desorption of the hydrogen adatoms.

In this work, we explore borophanes as hydrogen-based phonon-mediated superconductors, drawing inspiration from prior findings of high superconducting critical temperatures ($T_{c}$) in hydrogen-rich compounds \cite{h3s, lah10_1, lah10_2}. This idea traces back to Ashcroft's seminal theoretical analysis in 1968, which predicted high $T_{c}$ in pure, highly compressed hydrogen, due to strong phonon-mediated superconducting pairing resulting from its low mass and high Debye temperature \cite{ashcroft}. However, major practical challenges are posed by the immense pressure required for production of solid metallic hydrogen \cite{hp1,hp2}. This has led to an alternative push for hydrogen-rich superconductors that are stable under ambient (pressure) conditions. Placing hydrogen adatoms onto low-dimensional materials has yielded a similar promise in this regard, as demonstrated for monolayers of MgB$_2$ \cite{jonas1}, 2D metal carbides and nitrides (MXenes) \cite{jonas2}, and elemental gallium monolayers dubbed gallenene \cite{jonas3}. 

Earlier studies have reported that certain borophene phases do not show the expected enhanced $T_{c}$ upon hydrogenation \cite{absence1, absence2}. In contrast, our findings indicate that hydrogenation of $\beta_{12}$ borophene plays a pivotal role in enhancing both its stability and superconducting characteristics compared with its bare form. This discrepancy primarily arises from the selection of the electronic smearing parameter, which greatly influences the electronic properties and, consequently, the superconducting behavior of borophene. We found that using a smaller smearing parameter (0.0025 Ry) than the commonly used 0.02 Ry \cite{sup1,sup2,absence1,absence2} is crucial for obtaining accurate computational results (cf.\ Figure S8 in the SI for further details). 
Furthermore, by employing fully anisotropic Migdal-Eliashberg theory, we provide detailed insight into the origins of the material's superconducting behavior on the atomic scale. Additionally, we uncover several experimentally feasible strategies (strain, carrier doping) to further boost borophane's superconducting performance, leading to $T_c$ values up to 29 K.

\begin{figure}[t!]
    \centering
    \includegraphics[width=\textwidth]{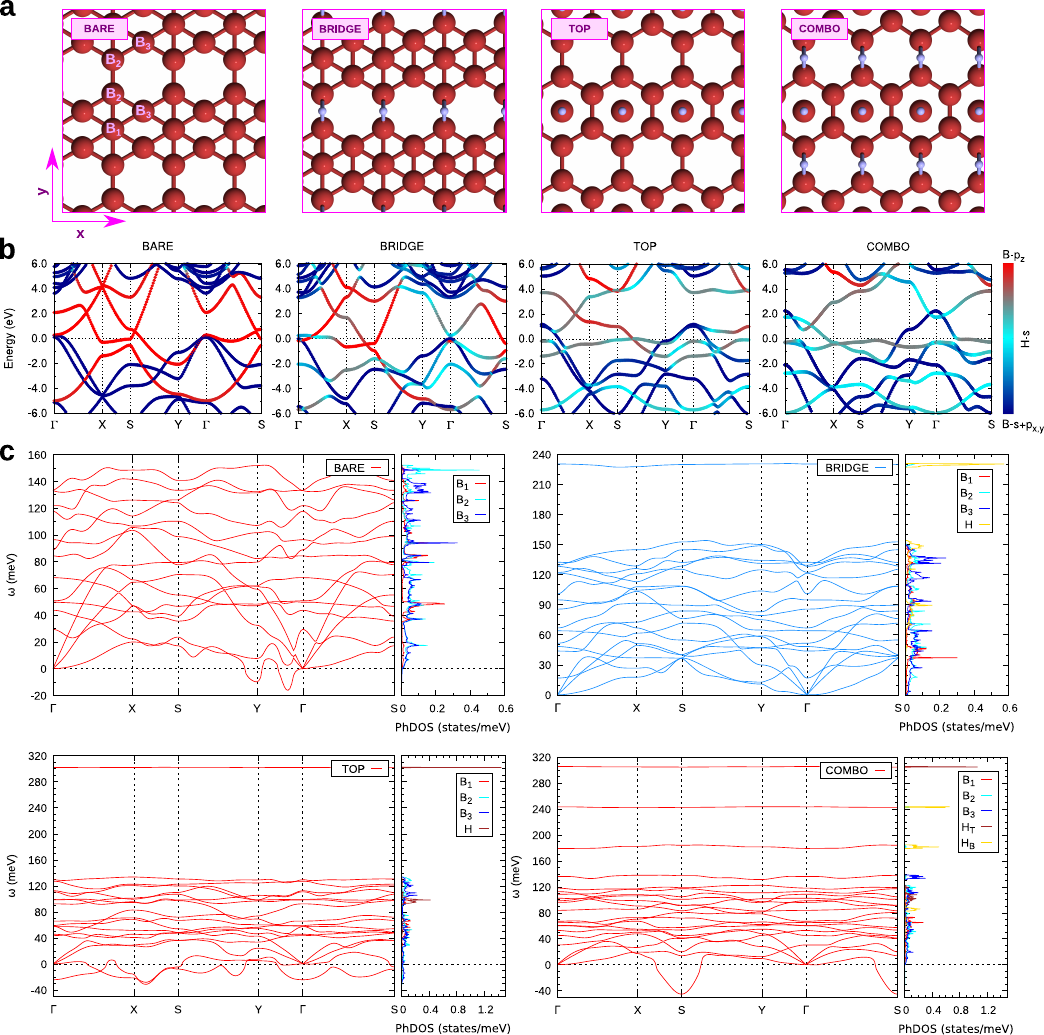}
    \caption{\textbf{Structural, electronic, and vibrational properties of bare and hydrogenated $\beta_{12}$ borophene.} (a) Top views of the crystal structures of bare and hydrogenated borophene (bridge, top, and combo adsorption site), together with their (b) orbital-resolved electronic band structures, and (c) corresponding phonon dispersions and atom-resolved phonon density of states (phDOS).}
    \label{fig:1}
\end{figure}

As the first step in our investigation, we determine the structural properties of bare and hydrogenated $\beta_{12}$ borophene, as shown in Figure \ref{fig:1}(a). The bare crystal structure consists of a planar sheet of boron (B) atoms arranged in a triangular lattice, with the exception of a row of hexagonal hollows with a missing boron atom at their center. The remaining occupied central B sites act as electron donors, while the hollows themselves are electron acceptors, creating a unique ``self-doping'' mechanism. Furthermore, the structure belongs to space group \textit{Pmmm} (No.\ 47), with different coordination numbers for the constituent boron atoms, influencing local electronic properties and their resulting reactivity, and enabling adsorption of hydrogen on different sites. Despite its inherent dynamical instability in the freestanding form, we achieve stabilization of $\beta_{12}$ borophene through hydrogen adsorption, pinpointing the optimal adsorption site at the ``bridge'' position (H$_{\mathrm{bridge}}$-$\beta_{12}$), characterized by the highest binding energy among all considered configurations, amounting to 3.39 eV. 

In Figure \ref{fig:1}(b), one observes that the bonding states ($\sigma$) of the in-plane B orbitals, composed of $s+p_{x,y}$ orbitals, are nearly fully occupied in the bare phase, with a gap of over 3 eV separating them from their antibonding counterparts ($\sigma^*$). The remaining electrons occupy the states composed of the out-of-plane B-$p_z$ orbitals. On the other hand, our Bader analysis reveals that hydrogen extracts 0.5$e$ from the borophene layer, thus altering the electronic band structure. Hydrogenation also changes the lattice symmetry, so that the H$_{\mathrm{bridge}}$-$\beta_{12}$ borophene structure corresponds to space group \textit{Pmm2} (No.\ 25). 

The phonon band structure of the bare $\beta_{12}$ borophene phase, obtained from our density functional perturbation theory (DFPT) calculations, displays a dynamical instability -- see Figure \ref{fig:1}(c). It arises mainly from the weak B$_2$--B$_2$ atom bond, resulting in unstable out-of-plane phonon modes, as displayed in Figure S5 in the Supporting Information (SI). However, when hydrogen is adsorbed at the ``bridge'' site, this bond is reinforced. As such, a robust B$_2$--H--B$_2$ three-center bond is formed, inducing slight buckling along with a minimal change in the in-plane lattice constants (with merely 0.34$\%$ tensile strain) in both directions. This strengthened bond is responsible for the dynamical stability of this hydrogenated compound (see Figure \ref{fig:1}(c)), in contrast with its bare electron-deficient form.   
 
\begin{figure}[ht!]
	\centering
	\includegraphics[width=\textwidth]{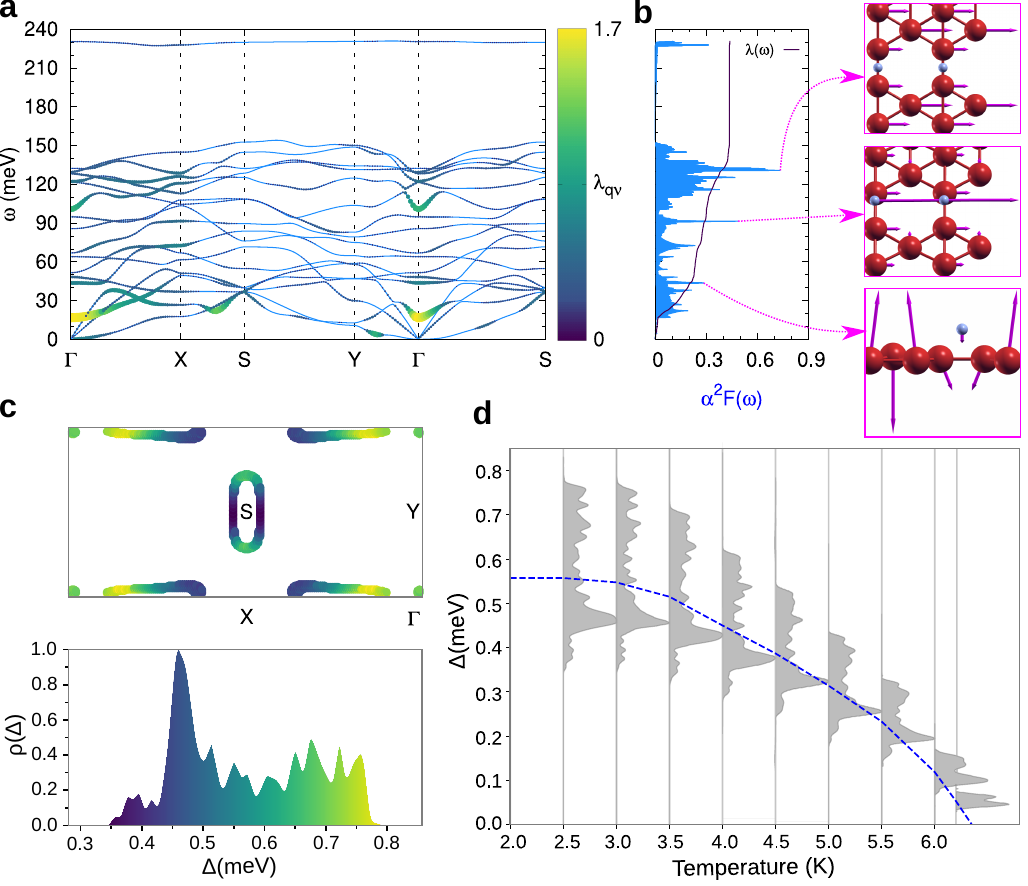}
	\caption{\textbf{Superconducting properties of H$_{\mathrm{bridge}}$-$\beta_{12}$ borophene, calculated with fully anisotropic Eliashberg approach.} (a) Phonon dispersion with the momentum-resolved electron-phonon coupling $\lambda_{\textbf{q}\nu}$ indicated by colors. (b) The isotropic Eliashberg function, and phonon modes corresponding to peaks in the Eliashberg function. (c) The superconducting gap spectrum on the Fermi surface, calculated at 2.5 K, and the corresponding distribution $\rho(\Delta)$ with the color code. (d) The evolution of $\rho(\Delta)$ with temperature, including weighted averages (dashed blue line), yielding $T_{c}^{\mathrm{aniso}}$ of 6.3~K.}
	\label{fig:2}
\end{figure}

The other chief hydrogenated configurations that have been experimentally identified in Ref.\ \citenum{borophane} include the top site, and a combination of the top and bridge sites which we refer to as ``combo'', both depicted in Figure \ref{fig:1}(a). Hydrogen on the top site of borophene leads to dynamical instability, exhibiting unstable phonon branches throughout the entire Brillouin zone (see Figure \ref{fig:1}(c)), accompanied by pronounced structural buckling. In contrast, the dynamical instability of the ``combo'' configuration is characterized by an unstable phonon mode localized around the $S$ point. It involves out-of-plane movement of the B$_2$ and H$_{\mathrm{bridge}}$ atoms along with in-plane movements of the B$_3$ and H$_{\mathrm{top}}$ atoms.

The electron deficiency of the ``combo'' configuration causes gapping between the B-$\pi$ and $\pi^{*}$ states (see Figure S13 in the SI). In view of these significant changes in the electronic structure, we have investigated the effect of applied carrier doping on dynamical stability. We found that adding electrons to this system so as to fill up more the $\pi^{*}$ states achieves dynamical stability, in analogy to the dynamically stable bridge configuration where the $\pi^{*}$ states are more occupied. Indeed, our simulations show that it takes adding 0.5 electrons to stabilize the structure (i.e., $0.1e$ per B atom), the amount which the additional hydrogen atom has extracted from the borophene layer. By shifting the Fermi level, a good electronic balance between the partially filled $\sigma$ and $\pi^{*}$ states can be achieved, which is crucial for dynamical stability of the material. However, the values of the isotropically evaluated $\lambda_{\mathrm{iso}}$ and $T_{c}^{\mathrm{iso}}$ remain small, with superconductivity even vanishing for doping levels beyond 0.1$e$/B (see Figure S15 in the SI). This is linked to the decreasing contributions of B$_1$-$p_z$, B$_2$-$p_z$ and H-$s$ states to the density of states at the Fermi level ($E_F$), while only contributions from B-$\sigma$ states remain significant (see Table S2 in the SI). Generally, when comparing these ``top'' and ``combo'' configurations with the H$_{\mathrm{bridge}}$-$\beta_{12}$ structure, they do not harbor notable superconducting properties. 

Let us now delve deeper into the vibrational and superconducting properties of the most promising candidate for superconductivity, namely the H$_{\mathrm{bridge}}$-$\beta_{12}$ borophene configuration. In Figure \ref{fig:2}(a), we present its phonon band structure alongside the mode($\nu$)- and momentum(\textbf{q})-dependent electron-phonon (\textit{e-ph}) coupling ($\lambda_{\mathbf{q}\nu}$). The strongest coupling is found around $\Gamma$ for phonon energy values of $\sim20$ meV. This corresponds to the softened $B_1$ optical phonon mode. In Figure \ref{fig:2}(b), the Eliashberg function reveals several significant peaks, with primary contributions stemming from two phonon modes. The first one corresponds to the in-plane and in-phase movement of B$_1$, B$_2$ and H (and slight in-plane and out-of-phase movement of B$_3$) at $\sim$90 meV. The second one -- the highest peak overall -- originates from the optical $A_u$ phonon mode, stemming from the in-plane movement of B atoms at $\sim$130 meV. The resulting overall isotropic \textit{e-ph} coupling constant amounts to $\lambda_{\mathrm{iso}} = 0.44$, yielding a critical temperature of $T_{c}^{\mathrm{iso}}$ = 4.0 K using the McMillan-Allen-Dynes (MAD) formula. 

To accurately assess the influence of hydrogenation on borophene's superconducting properties, we employ our first-principles results as input to solve the fully anisotropic Migdal-Eliashberg equations. Through this approach, we obtain the superconducting gap $\Delta(\mathbf{k})$ as a function of electron momentum $\mathbf{k}$ at the $E_F$, shown in Figure \ref{fig:2}(c). The Fermi surface is composed of three sheets: (i) an ellipsoidal Fermi sheet located around the $S$ point, with hybridized B-$p_{z}$ and H-$s$ character, (ii) an elongated sheet extending along $\Gamma$-$X$, where the character varies from purely B-$p_z$ (close to $\Gamma$) to hybridized B-$p_z$/H-$s$ further away from $\Gamma$, and (iii) a small circular sheet around $\Gamma$ with purely B-$\sigma$ character. The Fermi sheet around $S$ hosts the lowest $\Delta(\mathbf{k})$ values (0.35 meV at 0 K), where the electronic character is purely B-$p_{z}$, and slightly higher values where the character is hybrid B-$p_{z}$/H-$s$. For the sheet along $\Gamma$-$X$, the lowest gap values (down to 0.43 meV) stem from hybridized B-$p_z$ and H-$s$ states, while the highest ones (up to 0.78 meV) stem from pure B-$p_z$ states. 

Nevertheless, the gap distribution $\rho(\Delta)$ (see Figure \ref{fig:2}(c)) does not present separate domes, hence, its superconductivity is anisotropic single-gap in nature. Figure \ref{fig:2}(d) shows the temperature evolution of $\Delta(\mathbf{k})$, obtained from solving the anisotropic Eliashberg equations for each temperature. This yields $T_c^{\mathrm{aniso}}$ = 6.3 K, the temperature where gap vanishes. Furthermore, to assess the contributions of the different chemical elements in H$_{\mathrm{bridge}}$-$\beta_{12}$ borophene, we also investigated the isotope effect. To this end, we renormalized the phonon frequencies according to the isotope masses and subsequently solved the anisotropic Eliashberg equations again. We found a stronger B isotope effect ($\alpha_{\mathrm{B}} = 0.63$) compared with the H isotope effect ($\alpha_{\mathrm{H}} = 0.05$), underscoring hydrogen's pivotal role in enhancing stability in bare $\beta_{12}$ borophene while maintaining its intrinsic superconducting properties.

\begin{figure}[ht!]
    \centering
    \includegraphics[width=0.9\textwidth]{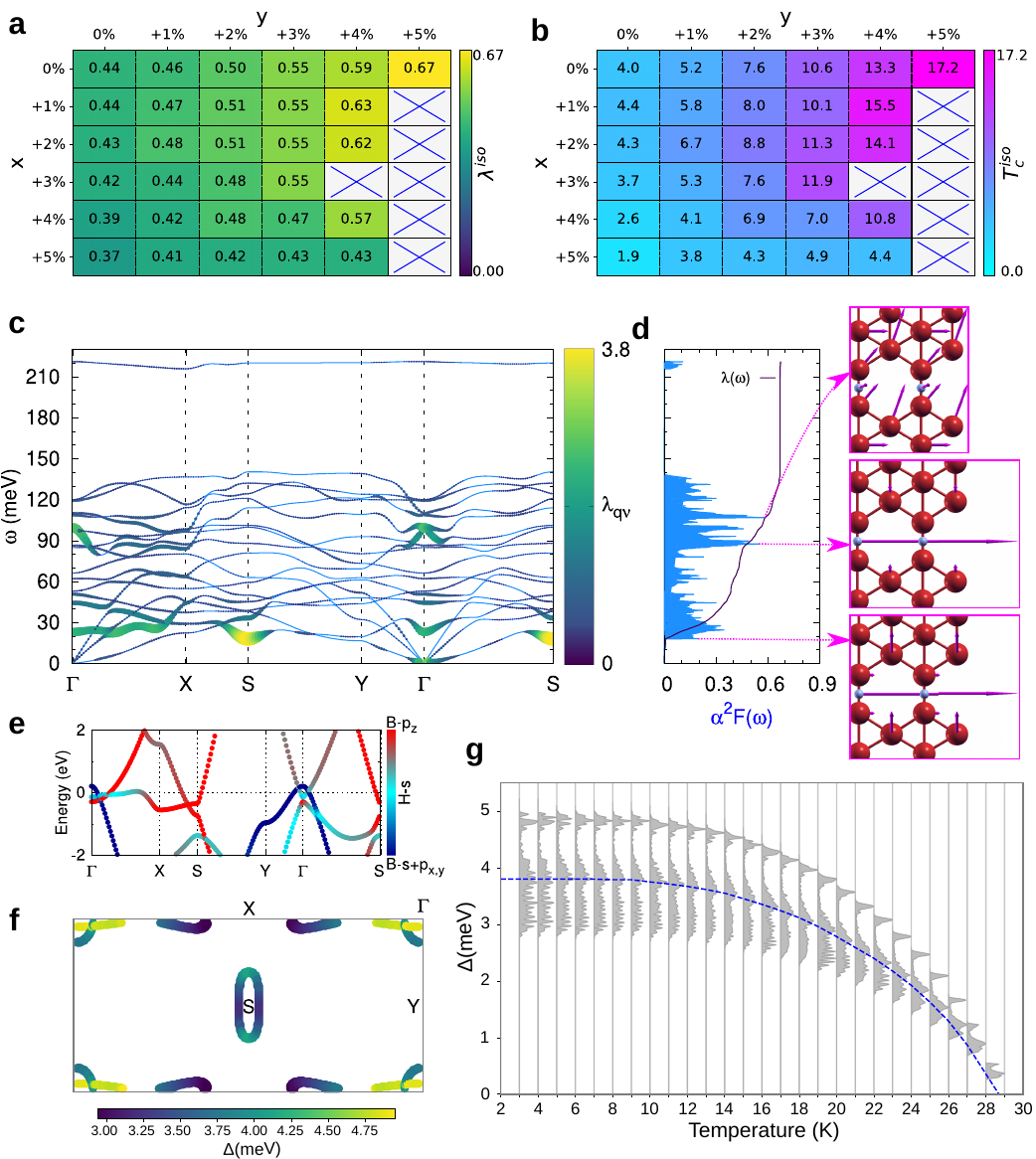}
    \caption{\textbf{Superconducting properties of H$_{\mathrm{bridge}}$-$\beta_{12}$ borophene under (anisotropic) in-plane tensile strain.} (a) The isotropic \textit{e-ph} coupling, and (b) corresponding $T_{c}^{\mathrm{iso}}$ values for considered tensile strains along the two principal in-plane directions, with maximal values found for for uniaxial tensile strain of $\epsilon_{y}=5\%$. For the latter optimal case, panel (c) shows the phonon dispersion with the momentum-resolved \textit{e-ph} coupling $\lambda_{\textbf{q}\nu}$ indicated by colors. (d) The corresponding isotropic Eliashberg function, and phonon modes corresponding to its selected peaks. (e) The corresponding orbital-resolved electronic band structure. (f) The superconducting gap spectrum on the Fermi surface, calculated at 3 K. (g) The evolution of $\rho(\Delta)$ with temperature, including weighted averages (dashed blue line), yielding $T_{c}^{\mathrm{aniso}} =  28.6$~K.}
    \label{fig:3}
\end{figure}

As a next step, we have considered the influence of lattice deformation on the superconducting properties of H$_{\mathrm{bridge}}$-$\beta_{12}$ borophene. Applying strain to 2D materials is experimentally viable through e.g.\ the selection of a suitable substrate, or using a piezoelectric substrate, and has proven to be a successful route to tailor and enhance superconductivity in various 2D superconducting systems \cite{PhysRevB.96.094510,jonas1}. Given the strong anisotropy of $\beta_{12}$ borophene, we applied the in-plane strain ($\epsilon$) not necessarily equal in the $x$ and $y$ directions (see Figure \ref{fig:1}(a)). We have considered values for $\epsilon_{x}$ and $\epsilon_{y}$ ranging from $-5\%$ (compressive) to $+5\%$ (tensile) strain, with a step size of $1\%$ -- yielding 121 strain combinations in total. 

Considering first the effect of strain on bare $\beta_{12}$ borophene, we found that it becomes dynamically stable under a biaxial tensile strain of $\epsilon_{x,y}=+3\%$, achieving a $T_{c}^{\mathrm{iso}}$ of 2.13 K, however, superconductivity is suppressed once the strain level reaches $\epsilon_{x,y}=+5\%$, due to a rapid decrease of the DOS at the $E_F$, contributed by the B-$\sigma$ states (see Table S1 and Figure S7 in SI).

Upon hydrogenation, we found that hydrogen significantly stiffens $\beta_{12}$ borophene due to its inherently high phonon frequencies, rendering it unable to sustain compressive strain. On the other hand, the phonon spectra remain stable for tensile strain up to $+5\%$ for the majority of strain combinations, as shown in Figure \ref{fig:3}(a,b). 

The figure reveals distinctive changes in $\lambda_{\mathrm{iso}}$ with applied strain, which can be attributed to the interplay of two effects. Firstly, in the absence of strain in the $y$ direction, as $\epsilon_{x}$ increases, the $p_z$ contributions from B$_1$ and B$_2$ atoms decrease, leading to a reduction of $\lambda_{\mathrm{iso}}$. Conversely, when no strain is applied in the $x$ direction, we observe an increasing contribution of $\sigma$ states from B$_1$ and B$_2$ atoms as $\epsilon_{y}$ increases, leading to a distinct enhancement of $\lambda_{\mathrm{iso}}$. Other strain combinations are governed by the interplay between these two elementary tendencies. Additionally, the contribution of H-$s$ states increases with strain regardless of the direction the strain is applied in (see Table S3 in SI). Furthermore, H$_{\mathrm{bridge}}$-$\beta_{12}$ borophene shows superior superconducting properties compared with its bare form under the same biaxial strain values (see Figure S7 in SI).

The above comprehensive analysis showed the H$_{\mathrm{bridge}}$-$\beta_{12}$ structure under a $5\%$ uniaxial tensile strain in the $y$ direction to be the best candidate for superconductivity. Our calculations for higher strain values show no significant further enhancement of the superconducting properties, indicating that saturation has been reached (see Figure S66 in SI). Its $\lambda_{\mathrm{iso}}$ reaches 0.67, with the corresponding highest $T_{c}^{\mathrm{iso}}$ of 17.2 K. The strongest $\lambda_{\textbf{q}\nu}$ value occurs at the $S$ point at low frequencies [see Figure \ref{fig:3}(c)]. The corresponding mode is composed of out-of-plane movement of B$_3$ atoms and in-plane movement of B$_2$ and H atoms [see inset of Figure \ref{fig:3}(d)]. The Eliashberg function in Figure \ref{fig:3}(d) exhibits two more significant peaks. The first one occurs at $\sim$90 meV stemming from in-plane H (and slight out-of-phase B$_3$) movement, and the highest $A$ mode at $\sim$110 meV, arising from in-plane B and slight out-of-plane H movement. 

As shown in Figure \ref{fig:3}(f), the Fermi surface exhibits strong anisotropy, prompting us to perform fully anisotropic Migdal-Eliashberg calculations to thoroughly characterize its superconducting properties. Within the ellipsoidal sheet located around the $S$ point, $\Delta(\textbf{k})$ originates from hybridized B-$p_z$ and H-$s$ states. Along $\Gamma$-$X$, the lower gap values stem from hybridized B-$p_z$ and H-$s$ states, while the highest values originate from the pure B-$p_z$ state. Despite some changes in the composition of the Fermi surface, these characters and how they relate to the $\Delta(\textbf{k})$ values are completely in line with the unstrained counterpart discussed before. Exhibiting a strongly anisotropic single-gap nature, the $\Delta(\textbf{k})$ values reach up to 4.95 meV at 0 K, corresponding to an enhanced $T_{c}^{\mathrm{aniso}}$ value of 28.6 K, as shown in Figure \ref{fig:3}(f). 

\begin{figure}[ht!]
	\centering
	\includegraphics[width=\textwidth]{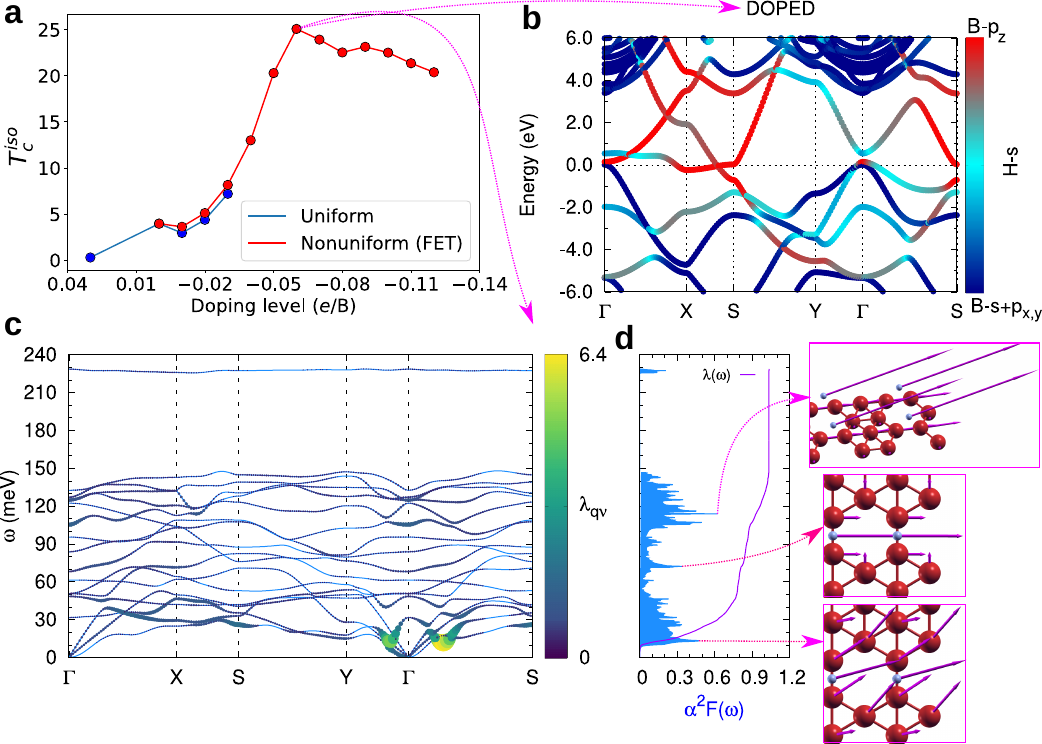}
	\caption{\textbf{Superconducting properties of H$_{\mathrm{bridge}}$-$\beta_{12}$ borophene under the influence of both uniform and non-uniform electron and hole doping.} (a) The dependence of $T_{c}^{\mathrm{iso}}$ on the doping level. The negative values correspond to hole doping as electrons are removed in this case. (b) The electronic band structure with the band character indicated by colors for 0.06 $e$/B hole doping. (c) The corresponding phonon dispersion with the momentum-resolved electron-phonon coupling $\lambda_{\textbf{q}\nu}$ indicated by colors. (d) The isotropic Eliashberg function, and phonon modes corresponding to the main peaks in the Eliashberg function.}
	\label{fig:4}
\end{figure}

Besides strain engineering, controlled charge doping is another established method to modulate the electronic properties of 2D materials, and can furthermore occur naturally as an additional effect of certain substrates. At the first level of approximation, we employed spatially uniform doping, where we observed only very minor changes in the lattice constants. This fact allowed us to subsequently apply non-uniform doping of the field-effect transistors (FET) type, within a single-sided gate geometry. Namely, this approach is feasible because, in a FET setup within Quantum ESPRESSO (QE), the structural relaxation cannot be performed directly. The applied electric field induces forces not accounted for in standard relaxation algorithms, and altering the lattice changes the doping charge per area and dipole effects, complicating the accurate energy calculations.

As the first step, we found that the bare form of $\beta_{12}$ borophene remains dynamically unstable upon both electron and hole doping. In contrast, for H$_{\mathrm{bridge}}$-$\beta_{12}$ borophene with a non-uniform hole doping level of 0.06 $e$/B, the structure is dynamically stable and $T_{c}^{\mathrm{iso}}$ reaches 25.1 K, as shown in Figure \ref{fig:4}(a).\footnote{We note that using the basic acoustic sum rule implemented in Quantum ESPRESSO (QE), the structures remain dynamically stable only up to a hole doping level of 0.03 $e$/B. Applying the general invariance and equilibrium conditions of the lattice potential, which recover the quadratic dispersion of flexural phonons in low-dimensional materials \cite{marzari}, the non-uniformly doped structures show dynamical stability up to 0.12 $e$/B hole doping level. As this dedicated acoustic sum rule is not implemented in the Electron-Phonon Wannier (EPW) package, we were restricted to obtaining the $T_c$ on the isotropic level in the subsequent analysis, using the MAD formula.}

Since the electronic DOS at the $E_F$ strongly affects $\lambda_{\mathrm{iso}}$, and is readily altered by carrier doping, we investigated its effect on the superconducting properties. We found that the DOS at $E_F$ decreases (increases) with electron (hole) doping (see Table S4 and S5 in SI). Notably, the contribution of B-$\sigma$ and B-$p_z$ states increases upon hole doping, boosting $\lambda_{\mathrm{iso}}$ up to 1.04 (at 0.06 $e$/B). The largest $\lambda_{\textbf{q}\nu}$ values stem from the softened $A$ phonon mode along $\Gamma$-$X$ at $\sim$10 meV, corresponding to in-plane movement of H atoms and rotating movement of B atoms [see Figure \ref{fig:4}(c,d)]. The other two important peaks at higher frequencies in the Eliashberg function are due to the $A_g$ phonon mode at $\sim$70 meV, stemming from in-plane B$_2$ and H (and slight out-of-phase B$_3$) movement, and the phonon mode at $\sim$110 meV corresponding to in-plane and in-phase movement of B$_2$ (slight in-plane and out-of-phase B$_3$) and out-of-plane movement of H atoms, as shown in Figure \ref{fig:4}(d). Overall, our findings indicate that hole doping is an efficient mechanism to enhance $\lambda_{\mathrm{iso}}$, next to the tensile strain. However, when applied simultaneously, the combined effect of strain and doping reduces $\lambda_{\mathrm{iso}}$ and $T_{c}^{\mathrm{iso}}$ due to the decreased contribution of B-$\sigma$ states at $E_{F}$ (see Figure S83 in SI). 

In conclusion, through precise calculations of the electronic and vibrational structures (paying due attention to the critical role of a sufficiently low electronic smearing), our study has revealed that hydrogen adsorption at the bridge site of $\beta_{12}$ borophene not only ensures dynamical stability of the material but also preserves its intrinsic superconducting properties, characterized by a $T_c$ of 6.3 K, as obtained from anisotropic Migdal-Eliashberg calculations. Moreover, this borophane system is prone to strong enhancement of the $T_c$ through tensile strain or hole doping, in contrast to the bare borophene phase, and stands out among other borophane phases for its promising superconducting characteristics. As a concrete example, we showed that applying uniaxial tensile strain of $5\%$ along the hydrogenated bridges increases the $T_{c}$ to 28.6 K in hydrogenated $\beta_{12}$ borophene, above the liquid hydrogen temperature (20.3 K), making future practical applications more viable. This increase goes hand in hand with the increasing contribution at the $E_F$ of the B-$\sigma$ and the B-$p_z$ states, with a minor contribution from the hybridized B-$p_z$/H-$s$ state. This further highlights the crucial role of both B-$\sigma$ and B-$p_z$ states in governing superconductivity, contrasting the bare phase where $\sigma$ states are predominant. We have also shown that similar values of $T_{c}$ in the vicinity of 30 K can be achieved by moderate hole doping, but this effect is not additive to the one of the strain. With such robust superconductivity and electronic and mechanical sensitivity, while also being more stable in ambient environments than its bare counterpart, our findings promote hydrogenated $\beta_{12}$ borophene as a highly promising platform for further development of boron-based 2D superconducting technologies.

\begin{acknowledgement}
This work was supported by the Research Foundation-Flanders (FWO). B.N.Š. acknowledges support from the Montenegrin Ministry of Science and the Special Research Fund of the University of Antwerp (BOF-UA) -- No.\ 542300011. J.B. acknowledges support as a Senior Postdoctoral Fellow of Research Foundation-Flanders (FWO) under Fellowship No.\ 12ZZ323N. The computational resources and services were provided by the VSC (Flemish Supercomputer Center), funded by the FWO and the Flemish Government -- department EWI. The collaborative effort in this work was fostered by EU-COST actions NANOCOHYBRI (Grant No.\ CA16218), SUPERQUMAP (Grant No.\ CA21144) and Hi-SCALE (Grant No.\ CA19108). The authors also thank D. Šabani and T. Pandey for fruitful discussions.

\end{acknowledgement}

\begin{suppinfo}

In the Supporting Information (SI) we provide computational details, as well as our complete set of results on the electronic, vibrational, and superconducting properties
of all considered compounds.
\end{suppinfo}

\bibliography{borophane.bib}

\end{document}